# Exploring refractory organics in extraterrestrial particles

Alexey Potapov[1], Maria Elisabetta Palumbo[2], Zelia Dionnet[3], Andrea Longobardo[4,5], Cornelia Jäger[1], Giuseppe Baratta[2], Alessandra Rotundi[4,5], Thomas Henning[6]

[1]*Laboratory Astrophysics Group of the Max Planck Institute for Astronomy at the Friedrich Schiller University Jena, Institute of Solid State Physics, Helmholtzweg 3, 07743 Jena, Germany, e-mail:* alexey.potapov@uni-jena.de

[2]*INAF-Osservatorio Astrofisico di Catania, Via Santa Sofia 78, 95123 Catania, Italy*

[3]*Université Paris-Saclay, CNRS, Institut d'Astrophysique Spatiale, 91405, Orsay, France*

[4]*Dip. Scienze e Tecnologie, Università degli Studi di Napoli "Parthenope," Centro Direzionale I C4, 80143 Napoli, Italy*

[5]*INAF-Istituto di Astrofisica e Planetologia Spaziali, Via Fosso del Cavaliere 100, 00133 Roma, Italy*

[6]*Max Planck Institute for Astronomy, Königstuhl 17, D-69117 Heidelberg, Germany*

**Abstract**

The origin of organic compounds detected in meteorites and comets, some of which could serve as precursors of life on Earth, still remains an open question. The aim of the present study is to make one more step in revealing the nature and composition of organic materials of extraterrestrial particles by comparing infrared spectra of laboratory-made refractory organic residues to the spectra of cometary particles returned by the Stardust mission, interplanetary dust particles, and meteorites. Our results reinforce the idea of a pathway for the formation of refractory organics through energetic and thermal processing of molecular ices in the solar nebula. There is also the possibility that some of the organic material formed already in the parental molecular cloud before it entered the solar nebula. The majority of the IR "organic" bands of the studied extraterrestrial particles can be reproduced in the spectra of the laboratory organic residues. We confirm the detection of water, nitriles, hydrocarbons, and carbonates in extraterrestrial particles and link it to the formation location of the particles in the outer regions of the solar nebula. To clarify the genesis of the species, high-sensitivity observations in combination with laboratory measurements like those presented in this paper are needed. Thus, this study presents one more piece of the puzzle of the origin of water and organic compounds on Earth and motivation for future collaborative laboratory and observational projects.

# 1. Introduction

One of the main hypotheses about the source of organic compounds that could serve as the basis of life on Earth is their formation in the parental molecular cloud or the protoplanetary disk and delivery to Earth through extraterrestrial objects (Oro 1961; Cronin & Chang 1993; Brack 1999; Pearce et al. 2017). Asteroids, interplanetary dust particles (IDPs), and, to some extent, comets bombarded intensively the surface of early Earth and could contribute directly with their organic content to the formation of a terrestrial prebiotic world (Chyba & Sagan 1992).

Organic compounds, which are critical for life as we know it, such as ribose, amino acids, and nucleobases, have been detected in meteorites and comets (Cronin & Chang 1993; Elsila et al. 2009; Cobb & Pudritz 2014; Altwegg et al. 2016) and synthesized in laboratory experiments simulating the evolution of icy dust grains in the interstellar medium (ISM) (Bernstein et al. 2002; Muñoz Caro et al. 2002; Nuevo et al. 2006; Elsila et al. 2007; Nuevo et al. 2012; Meinert et al. 2016; Oba et al. 2016; Krasnokutski et al. 2020) and circumstellar disks around young stars (Potapov et al. 2022). Moreover, meteorites, comets, and IDPs contain refractory organic compounds (Cooper et al. 2001; Matrajt et al. 2004; Keller et al. 2006; Rotundi et al. 2008; Dartois et al. 2013; Rotundi et al. 2014; Goesmann et al. 2015; Dartois et al. 2018; Dionnet et al. 2018). Such compounds may be reproduced in laboratory experiments by gas phase condensation of amorphous carbon-based materials (Colangeli et al. 1995; Hammer et al. 2000; Fanchini et al. 2002; Rodil 2005; Jäger et al. 2008; Quirico et al. 2008) and by UV or ion irradiation of molecular ices containing C-, H-, O-, and N-bearing molecules, such as $H_2O$, CO, $N_2$, $CO_2$, $NH_3$, $CH_4$, $CH_3OH$ (the most abundant ice species observed in astrophysical environments and comets) and their subsequent heating to room temperature leading to the formation of refractory organic residues (Munoz Caro & Schutte 2003; Palumbo et al. 2004; Fresneau et al. 2017; Accolla et al. 2018; Poch et al. 2020; Potapov et al. 2021).

The origin of the organic compounds detected in meteorites, comets and IDPs still remains an open question. The main issues are their amount, chemical pathways, formation conditions (including formation location in the solar nebula), and alteration by chemical processing within the Nebula. Answering these questions will help not only to reveal the astrochemical heritage of the Solar System but also to draw a more complete picture of the physico-chemical processes in interstellar and circumstellar media due to their common link (Ehrenfreund & Charnley 2000; Caselli & Ceccarelli 2012; Henning & Semenov 2013; Sandford et al. 2020).

In order to identify components of primitive bodies and dust grains in the Solar System and to constrain the origin and evolution of refractory organics inside extraterrestrial matter, it is critical to compare the analogs obtained in the laboratory with extraterrestrial samples collected on Earth, or directly in space thanks to sample return missions. Similarities in optical and structural properties and chemical compositions of extraterrestrial and laboratory-synthesized materials have been shown. However, a direct comparison of the spectra of real cosmic particles and their analogs is relatively rare. A perfect example of such a comparison is the assignment of the 3.2 μm feature in the spectrum of the comet 67P/Churyumov-Gerasimenko to ammonium salts produced in the laboratory (Poch et al. 2020). Ultracarbonaceous Antarctic micrometeorites exhibit IR signatures that are similar to those observed in the laboratory-synthesized carbon-based materials and organic residues (Dartois et al. 2013; Baratta et al. 2015; Accolla et al. 2018; Dartois et al. 2018). It was also suggested that interstellar ices should lead to organic materials enriched in heteroatoms that present similarities with cometary materials but strongly differ from meteoritic organic materials (Fresneau et al. 2017).

This study presents a comparison of the IR spectra of laboratory-made refractory organic residues produced by UV or ion irradiation of simple molecular ices and their subsequent heating to room temperature to the spectra of cometary particles returned by the Stardust mission (hereafter, Stardust particles), IDPs, and meteorites. Our aim is to make one more step in revealing the nature and composition of organic materials of extraterrestrial particles.

## 2. Methods
### 2.1. Extraterrestrial particles and their spectral measurements

In this section, we present the extraterrestrial samples chosen as a point of reference for the comparison with the analogs: (i) two meteoritic fragments, (ii) two IDPs, and (iii) three cometary particles collected by the Stardust mission.

We selected two very different samples of carbonaceous chondrites: (i) the Paris meteorite (petrologic type: CM2.7-2.8), as it is one of the less altered carbonaceous chondrites available in the laboratory (Hewins et al. 2014) and (ii) a fragment of the North West Africa (NWA) 5515 meteorite, a CK4 chondrite representative of meteorites, which has undergone thermal metamorphism. Our aim was to compare the laboratory samples with two meteoritic samples with very different types of alteration. The classification of meteorite types is presented elsewhere (Rubin et al. 2007). IR measurements of the Paris meteorite were performed in transmission on a fragment (50 x 40 x 3 μm$^3$) crushed inside diamond windows. An Agilent (model Cary 670/620) microspectrometer, with its internal source (Globar), and a 25x objective, installed on the SMIS (Spectroscopy and Microscopy in the Infrared using Synchrotron) beamline of the SOLEIL synchrotron (France), was used to obtain data in the spectral range 4000-800 cm$^{-1}$ with a resolution of 4 cm$^{-1}$. The sample's average spectrum is used here. A complete analysis of this sample is presented elsewhere (Dionnet et al. 2018). The spectrum of NWA 5515 was collected on a 3D grain with the same setup (Aléon-Toppani et al. 2021).

The spectra of IDPs were another point of comparison for the laboratory analogs. The sample L2021C5 has been crushed in diamond windows and its IR transmission spectrum was measured at the SMIS beamline of the SOLEIL synchrotron with a NicPlan microscope, coupled to a Magna 860 FTIR spectrometer (Thermo Fisher) operating in transmission in the 5000–650 $cm^{-1}$ range and with a resolution of 4 $cm^{-1}$. A complete analysis of the particle is presented elsewhere (Brunetto et al. 2011). The U217B19 IR spectrum has been presented and analyzed by Ishii et al. (2018). The IR data were provided by Hope Ishii.

Then, we compared analogs' spectra with those of Stardust particles directly collected in the coma of the comet 81P/Wild and brought back to Earth in 2006. Collected samples were trapped in aerogel. A few of them were analyzed thanks to IR spectroscopy (Rotundi et al. 2008), namely particles 35.17 (sized 10 x 7 $\mu m^2$), 35.21 (sized 20 x 13 $\mu m^2$) and 35.26 (sized 11 x 19 $\mu m^2$). Particles 35.17, 35.21, and 35.26 were deposited on a KBr window and then IR data have been collected. The particle 35.17 was studied in transmission by the LANDS team at the Laboratorio di Fisica Cosmica e Planetologia (LFCP), Napoli, with a microscope attached to an FTIR interferometer (Mod. Bruker Equinox-55) in the range 7000–600 $cm^{-1}$ and a spectral resolution of 4 $cm^{-1}$. On the opposite, the spectra of particles 35.21 and 35.26 have been measured in reflection by the Orsay-IAS group using a NicPlan microscope associated with the Magma 860 FT-IR spectrometer equipped with MCT detectors, in the range 4000–650 $cm^{-1}$ and with a spectral resolution of 4 $cm^{-1}$. Correction of the Stardust particle's spectra was applied by subtraction of the normalized spectra of the aerogel, measured in the surrounding pixels.

Concerning the Stardust samples, the organic contribution of the aerogel has been removed from the spectra. For each particle, a spectrum of the surrounding aerogel was taken and then subtracted. Water is more of a challenge. The samples could contain small amounts of adsorbed water due to air measurements and conservation of the samples. Concerning IDPs, some of them were sent inside silicone oil, but they were cleaned with hexane which should remove the contamination. It is possible that there are small remains for bands around 1281–1257 and 1261–1259 $cm^{-1}$ (7.81-7.96 and 7.93-7.94 μm). Many analyses have been done on the organic matter of IDPs with such a protocol, for instance, (Merouane et al. 2014).

### 2.2. Experiments in Jena

Residues formed after UV irradiation of simple ices deposited onto silicate grains were obtained in the Laboratory Astrophysics Group of the Max Planck Institute for Astronomy (Germany).

The experimental setup and procedure have been presented in two recent papers devoted to dust/ice mixtures (Potapov et al. 2018a; Potapov et al. 2018b). In brief, nanometre-sized amorphous $MgSiO_3$ silicate grains were produced by laser ablation of a Mg:Si (1:1) target in a quenching atmosphere of 4 Torr $O_2$. After the formation, grains were extracted through a nozzle and a skimmer from the ablation chamber and deposited on KBr substrates at the temperature of 10 K and the pressure in the deposition chamber of $5\times10^{-8}$ mbar. Grains interact on the substrate forming a porous layer of fractal aggregates with sizes of up to several tens of nanometres (Jäger et al. 2008; Sabri et al. 2014). The thickness of the grain deposits was controlled by a quartz crystal resonator microbalance using known values for the deposit area of 1 cm$^2$ and density of 2.5 g cm$^{-3}$ and was about 150 nm.

Ice mixtures $CH_3OH:H_2O$, $CH_3OH:H_2O:NH_3$, $CO_2:H_2O:NH_3$ and $CH_4:H_2O:NH_3$ containing C-, N- and H-bearing ice molecules abundant in astrophysical environments (Boogert et al. 2015) with volume ratios of about 1 for $MgSiO_3$/C-bearing molecule and 10 for C-bearing molecule/$H_2O$ and C-bearing molecule/$NH_3$ were deposited on silicate grains. The initial molecular compositions of the ices were chosen to address the efficient formation of refractory organic residues. The $H_2O$, $CO_2$, $CH_4$, $NH_3$, and $CH_3OH$ ice thicknesses were calculated from their vibrational bands at 3250, 2342, 3009, 1073 cm$^{-1}$, and 1026 cm$^{-1}$ using the band strengths of $2.0\times10^{-16}$ (Gerakines et al. 1995), $7.6\times10^{-17}$ (Gerakines et al. 1995), $5.7\times10^{-18}$ (Hudgins et al. 1993), $1.7\times10^{-17}$ (d'Hendecourt & Allamandola 1986), and $1.8\times10^{-17}$ cm molecule$^{-1}$ (Hudgins et al. 1993) correspondingly. The deposition time in all experiments was about 15 minutes.

After deposition, the silicate/ice mixtures were irradiated for 6 hours by a broadband deuterium lamp (L11798, Hamamatsu) with a flux of $10^{15}$ photons cm$^{-2}$ s$^{-1}$. Thus, the final fluence was about $2\times10^{19}$ photons cm$^{-2}$. The lamp has a broad spectrum from 400 to 118 nm with the main peaks at 160 nm (7.7 eV) and at about 122 nm (10.2 eV) corresponding to the emission of molecular and atomic hydrogen. After irradiation, the cooling was stopped and the samples were warmed up to room temperature. IR spectra of the samples at room temperature were taken *in situ* using an FTIR spectrometer (Vertex 80v, Bruker) in the transmission mode at a resolution of 1 cm$^{-1}$. The spectra of pure KBr substrate recorded at room temperature before the depositions and irradiation experiments were used as reference spectra.

### 2.3. Experiments in Catania

Residues formed after ion irradiation of simple ices were obtained in the Laboratory for Experimental Astrophysics at INAF-Osservatorio Astrofisico di Catania (Italy). The spectra of the residues have been already presented in a previous study (Accolla et al. 2018).

Gas mixtures were prepared in a mixing chamber and admitted into the UHV chamber (P < $10^{-9}$ mbar) through a needle valve. Inside the chamber, the gas condenses on a cold KBr substrate (17 K) in thermal contact with the cold finger of a closed-cycle He cryostat.

The UHV chamber is interfaced with a 200 kV ion implanter (Danfysik 1080-200). The thickness of the ice samples was measured during accretion by a laser interference technique. After deposition, each ice sample was irradiated with 200 keV $H^+$. After ion irradiation, the sample was warmed up to room temperature.

During irradiation, the ion current is kept below 1 µA/cm$^2$. The dose (in units of eV/16u, where u is the unified atomic mass unit defined as 1/12 of the mass of an isolated atom of carbon-12) is obtained from the knowledge of the fluence (that is measured during the experiment; in units of ions/cm$^2$) and the stopping power (that is obtained by SRIM software in units of eV cm$^2$ /molecules) being the thickness of the ice sample lower than the penetration depth of 200 keV protons. For the spectra presented in this work the dose was 120 eV/16u ($CH_4$:CO = 1:1), 126 eV/16u ($N_2$:$CH_4$ = 1:1), 106 eV/16u ($N_2$:$CH_4$:CO = 1:1:1), and 122 eV/16u ($N_2$:$CH_4$:$H_2O$ = 1:1:1) respectively. The initial ice molecular composition was chosen to optimize the formation of the corresponding refractory organic residue.

The vacuum chamber is placed in the sample compartment of an FTIR spectrometer (Vertex 70, Bruker). Transmission IR spectra are taken through KBr windows in the chamber and a hole in the cold finger before ice deposition (background spectrum), after deposition, after each step of ion irradiation, at selected temperature values during warm-up and at room temperature. Spectra of the residues shown in this work were taken at room temperature, the day after the irradiation, without breaking the vacuum. Spectra are taken at a resolution of 1 cm$^{-1}$ and sampling 0.25 cm$^{-1}$. The sample holder forms an angle of 45 degrees with the IR beam and the ion beam. As a consequence, IR transmission spectra can be taken at each experimental step without rotating the sample. More details on the experimental setup and procedure can be found elsewhere (Urso et al. 2016; Accolla et al. 2018).

### 3. Results

VUV photons or ions cause carbonization processes of the ices that lead to the formation of refractory materials. Depending on the irradiation dose, these materials can be partly or completely refractory corresponding to partly soluble or insoluble organic matter. The best analog of the samples produced in our experiments is kerogen. The nature of kerogen is not well defined. Its disordered structure can be described as a mixture of aromatic and aliphatic sub-units containing a large number of functional groups. It is now well accepted that the refractory organic materials in cometary dust as well as in meteorites are dominated by high molecular weight organic components very similar to the kerogen-like material (Osawa et al. 2009; Matthewman et al. 2013; Wooden et al. 2017).

The IR absorption spectra of the samples obtained after UV or ion irradiation of molecular ices and subsequent heating to room temperature are presented in Figure 1.

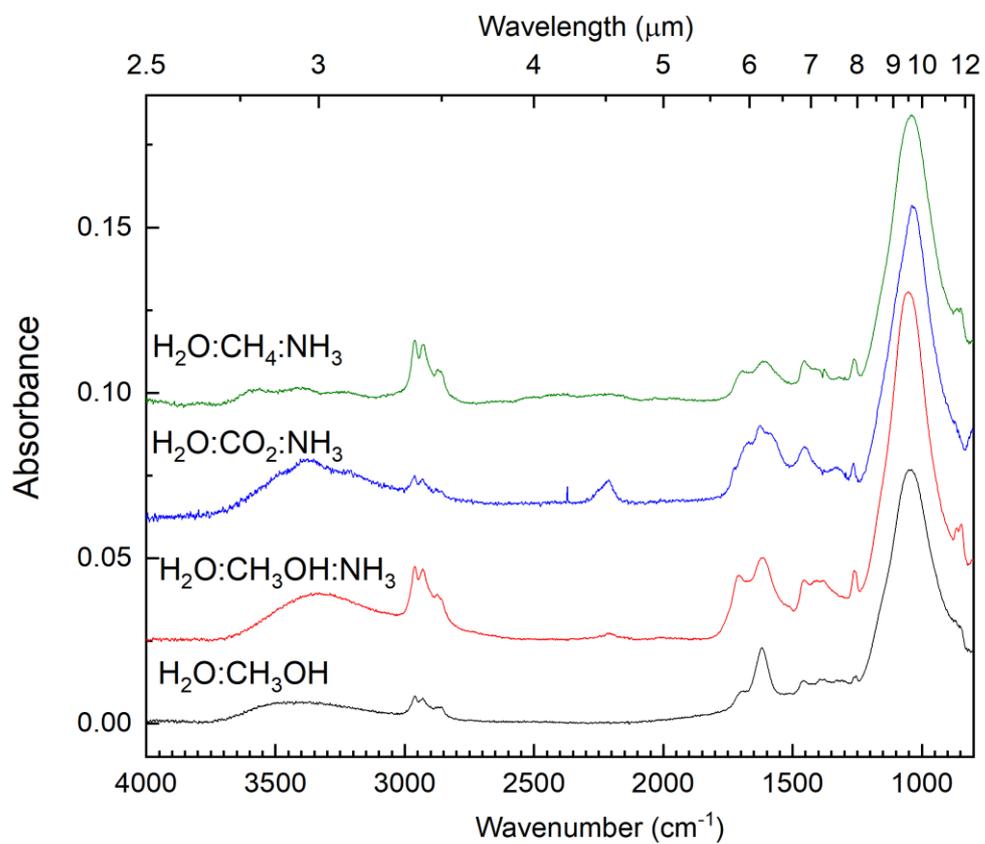

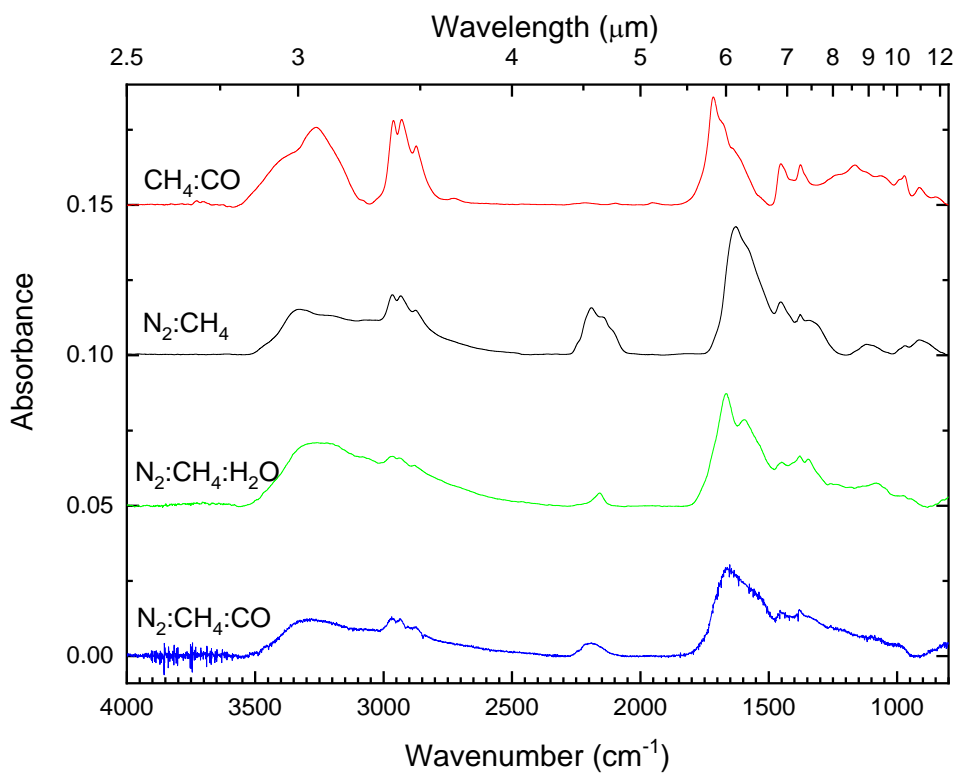

Figure 1. The IR absorption spectra of the laboratory organic residues. Upper - produced in Jena on the surface of silicate grains by UV irradiation of different ice mixtures (mentioned in the Figure) at 10 K and subsequent heating of the samples up to room temperature. Lower - produced in Catania on KBr substrates by ion irradiation of different ice mixtures (mentioned in the Figure) at 17 K and subsequent heating of the samples up to room temperature. The spectra are vertically shifted for clarity.

Comparison of low-temperature spectra before and after energetic processing is out of the scope of the present study. Examples of such a comparison can be found in the literature (Islam et al. 2014; Munoz Caro et al. 2019). The ice mixtures used for the production of laboratory organic residues, the detected IR bands of the residues and their attributions are listed in Tables 1 and 2.

Table 1. IR bands (cm$^{-1}$) of the Jena samples and their attributions.

| $CH_3OH:H_2O$ | $CH_3OH:H_2O:NH_3$ | $CO_2:H_2O:NH_3$ | $CH_4:H_2O:NH_3$ | Band attributions |
|---|---|---|---|---|
| 3750 – 3000 | 3750 – 3000 | 3750 – 3000 | 3750 – 3000 | OH stretching, NH stretching(Munoz Caro & Schutte 2003) |
| 2962, 2932 | 2962, 2932 | 2962, 2932 | 2962, 2930 | CH stretching (asym)(Colangeli et al. 1995) |
| 2876, 2858 | 2875, 2859 | 2879, 2859 | 2872, 2858 | CH stretching (sym)(Colangeli et al. 1995) |
| | | | 2523 | ? |
| | | | 2410 | ? |

|  |  | 2371, 2365 |  | ? |
|  | 2260-2175 | 2290-2155 | 2325-2096 | C≡N stretching(Accolla et al. 2018) |
| 1697, 1618 | 1707, 1615 | 1715, 1622 | 1697, 1610 | C=O, C=C, C=N stretching(Munoz Caro & Schutte 2003) HOH bending, $NH_2$ bending(Accolla et al. 2018) |
| 1513 | 1512 | 1516 | 1515 | $CO_3^{2-}$ stretching(Aguiar et al. 2009; Ciaravella et al. 2018) |
| 1458 | 1458 | 1455 | 1455 | CH bending(Munoz Caro & Schutte 2003) |
| 1427 |  |  | 1432 | $CO_3^{2-}$ stretching(Aguiar et al. 2009; Ciaravella et al. 2018) |
| 1394, 1377 | 1380 |  |  | CH bending (sym)(Colangeli et al. 1995; Jäger et al. 2008) |
| 1330 | 1330 | 1334 | 1320 | $COO^-$ stretching (sym)(Munoz Caro & Schutte 2003) |
| 1300 |  |  | 1300 | ? |
| 1263, 1255 | 1263, 1257 | 1265 | 1263, 1258 | ? |
| 1040 | 1040 | 1040 | 1040 | SiO stretching |
| 860 | 865, 847 |  | 865, 850 | out-of-plane CH bending(Colangeli et al. 1995; Jäger et al. 2008) |
|  | 804 | 804 | 804 | ? |

Table 2. IR bands (cm$^{-1}$) of the Catania samples and their attributions.

| **CH$_4$:CO** | **N$_2$:CH$_4$** | **N$_2$:CH$_4$:CO** | **N$_2$:CH$_4$:H$_2$O** | **Band attributions** |
|---|---|---|---|---|
| 3500 – 3000 | 3500 – 3000 | 3500 – 3000 | 3500 – 3000 | OH stretching, NH stretching(Munoz Caro & Schutte 2003) |
| 2961, 2930 | 2966, 2933 | 2966, 2934 | 2966, 2934 | CH stretching (asym)(Colangeli et al. 1995) |
| 2873 | 2873 | 2873 | 2877 | CH stretching (sym)(Colangeli et al. 1995) |
|  | 2270 - 2050 | 2270 - 2050 | 2270 - 2050 | C≡N stretching(Accolla et al. 2018) |
| 1716, 1628 | 1628, 1580 | 1662 | 1666, 1660 | C=O, C=C, C=N stretching(Munoz Caro & Schutte 2003) HOH bending, $NH_2$ bending(Accolla et al. 2018) |
| 1454 | 1453 | 1452 | 1450 | CH bending(Munoz Caro & Schutte 2003) |
| 1376 | 1378 | 1379 | 1380 | CH deform. (sym)(Colangeli et al. 1995; Jäger et al. 2008) |
|  | 1320 | 1340 | 1348 | $COO^-$ stretching (sym)(Munoz Caro & Schutte 2003) |
| 1240 |  | 1260 | 1240 | ? |
| 1161 |  |  |  | ? |
|  | 1100 | 1095 | 1085 | ? |
| 1057 |  |  |  | ? |
| 990 |  | 990 |  | ? |
| 970 | 970 |  | 970 | ? |
|  |  |  | 940 | ? |
| 911 | 911 |  |  | ? |
| 850 |  |  |  | out-of-plane CH bending(Colangeli et al. 1995; Jäger et al. 2008) |

All residue spectra show similar profiles containing bands that can be attributed to vibration modes of different refractory organic materials (additionally, the Si-O stretching band around 1000 cm$^{-1}$ in Jena samples containing amorphous MgSiO$_3$ grains). When an N-bearing species ($N_2$ or $NH_3$) is present in the original ice mixture, a broad feature at about 2200 cm$^{-1}$ (4.54 μm) is observed in the spectra of the residues (see Figure 1 and Tables 1 and 2). This feature is assigned to nitriles and isonitriles (C≡N triple bond) (Accolla et al. 2018). From a qualitative point of view, the spectra of the laboratory samples are very similar to each other. The similarity is due to the fact that the organic residues have the same functional groups, such as OH and NH stretching vibrations, $CH_3$, $CH_2$, and C=O, C=C, C=N, stretching vibrations, N-H and CH bending vibrations, and so on. The majority of the observed IR features have the same or slightly shifted band positions (see Tables 1 and 2). The spectral bands not present in both data sets are the following:

(i) The silicate band at 1040 cm$^{-1}$ (9.62 μm) in the Jena spectra due to the presence of $MgSiO_3$ grains.

(ii) The weak bands at about 1515 and 1430 cm$^{-1}$ (6.58 and 6.99 μm) in the Jena spectra attributed to the $CO_3^{2-}$ stretching in magnesium carbonates ($MgCO_3$).

(iii) The bands at 1628 and 1580 cm$^{-1}$ (6.14 and 6.33 μm) in the Catania spectra, which can be attributed to C=C and/or C=N stretching. In combination with the more intense and broad C≡N 2200 cm$^{-1}$ band (as compared to the Jena samples), this result points to the more efficient formation of N-bearing residues by ion irradiation of $N_2$-containing ices.

(iv) A number of bands without attributions in both spectral sets.

(v) In addition, the bands ranging from about 1650 to 1720 cm$^{-1}$ (6.06 to 5.81 μm) in the Catania samples are broader and show different shapes compared to the Jena samples. This might point to the presence of N-H bending vibrations in addition to C=O or C=N groups caused by the presence of -$NH_2$ groups in the ion-irradiated samples (Bernstein et al. 1995; Krasnokutski et al. 2022; Oba et al. 2022). However, a clear identification of the functional groups and compositions of the processed organics based on the IR spectra is difficult. Additional analytical characterization methods such as mass spectrometry, X-ray Photoelectron Spectroscopy (XPS), and X-ray absorption near-edge structure (XANES) are required.

A detailed study on the reproducibility and stability of the residues produced after ion irradiation of icy samples in Catania has been reported (Baratta et al. 2015; Accolla et al. 2018; Baratta et al. 2019). In particular, 30 residues have been produced after ion irradiation of icy samples made of $N_2$:$CH_4$:CO=1:1:1 deposited at 17 K (Baratta et al. 2015). Icy samples of three different thicknesses have been irradiated at the same ion dose (110 ± 5 eV/16u). It has been shown that the thickness of the final residue is proportional to the thickness of the initial ice sample and that the profile and relative intensity of IR bands do not depend on the sample thickness. Furthermore, the stability of the residues has been checked over a period of about 1200 days (Baratta et al. 2019). It has been found that the intensity of the -C≡N feature at about 2190 cm$^{-1}$ (4.57 µm) decreased at the beginning and then stabilized after 200 days and did not vary further, within uncertainties, after 700 days. No significant variations have been observed for the other IR bands.

The experiments on the UV irradiation of ices did not produce completely reproducible results. The ice mixtures produced in Jena were irradiated with VUV photons for 6 h. Under these conditions, the carbonization process of the ices was well advanced leading to the formation of a kerogen-like material. This has been proven by ex-situ analysis of the residues (Jäger et al., in preparation). The analyses performed by high-resolution transmission electron microscopy (HRTEM), energy dispersive X-ray (EDX) spectroscopy, and electron energy loss spectroscopy (EELS) have shown that the content of C, O, and N can vary in a nanometer scale within one sample. In our samples, a number of functional groups such as ≡CH, =CH$_2$, -CH$_3$, C=O, HC=O, -COOH, NHx, as well as ether and ester groups are formed. However, the ratios of these functional groups can differ as well as the carbon structure in a nanometer scale, but the general trend of the formation of O- and N-bearing groups is the same. We have to note that IR spectroscopy is not sensitive enough to detect small changes in the kerogen-like carbon structure.

In the following, we present the results of the comparison of the IR spectra of meteorites (primitive carbonaceous chondrites), IDPs, and Stardust particles to the spectra of organic residues produced in the laboratory. The spectra of the residues produced from N-free and N-containing ice mixtures are quite similar except the C≡N feature around 2200 cm$^{-1}$ (4.54 µm), that is why only the spectra for the N-containing mixtures are shown in the next three Figures.

We compared the laboratory spectra with the spectra of meteoritic samples, namely the Paris meteorite and the NWA 5515 meteorite. Figure 2 shows the comparison of the absorption spectra of the meteorites and laboratory organic residues. The detected IR bands and their attributions for the meteorites are presented in Table 3.

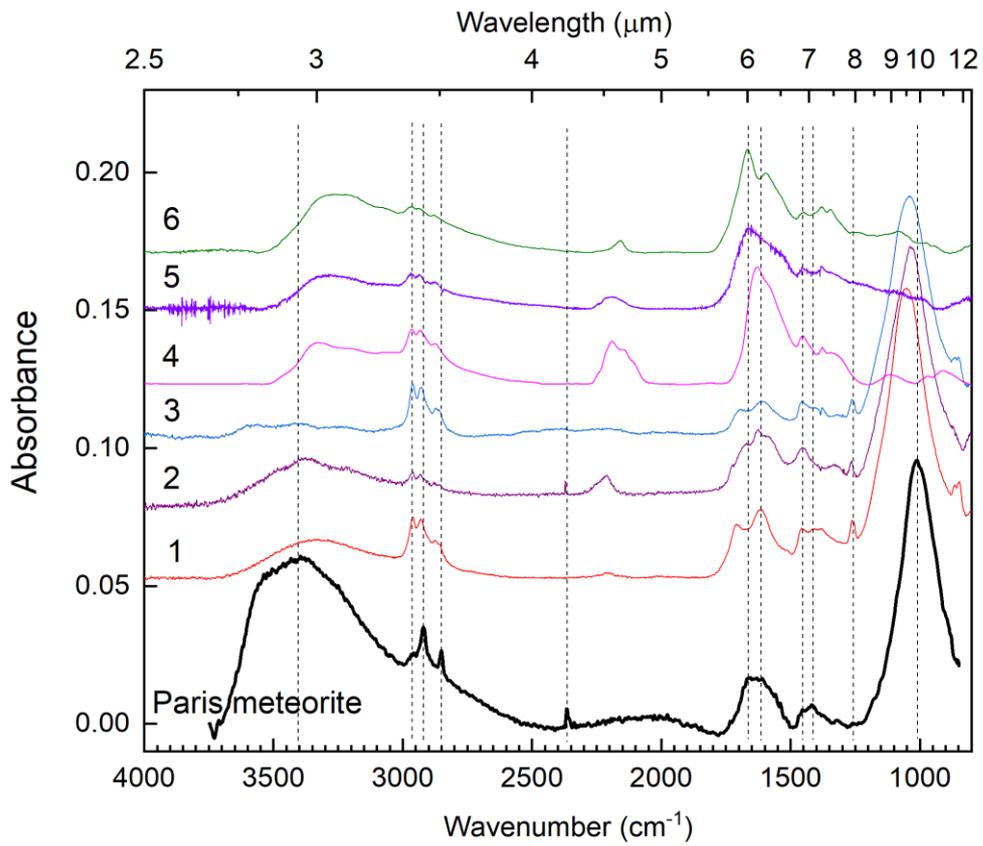

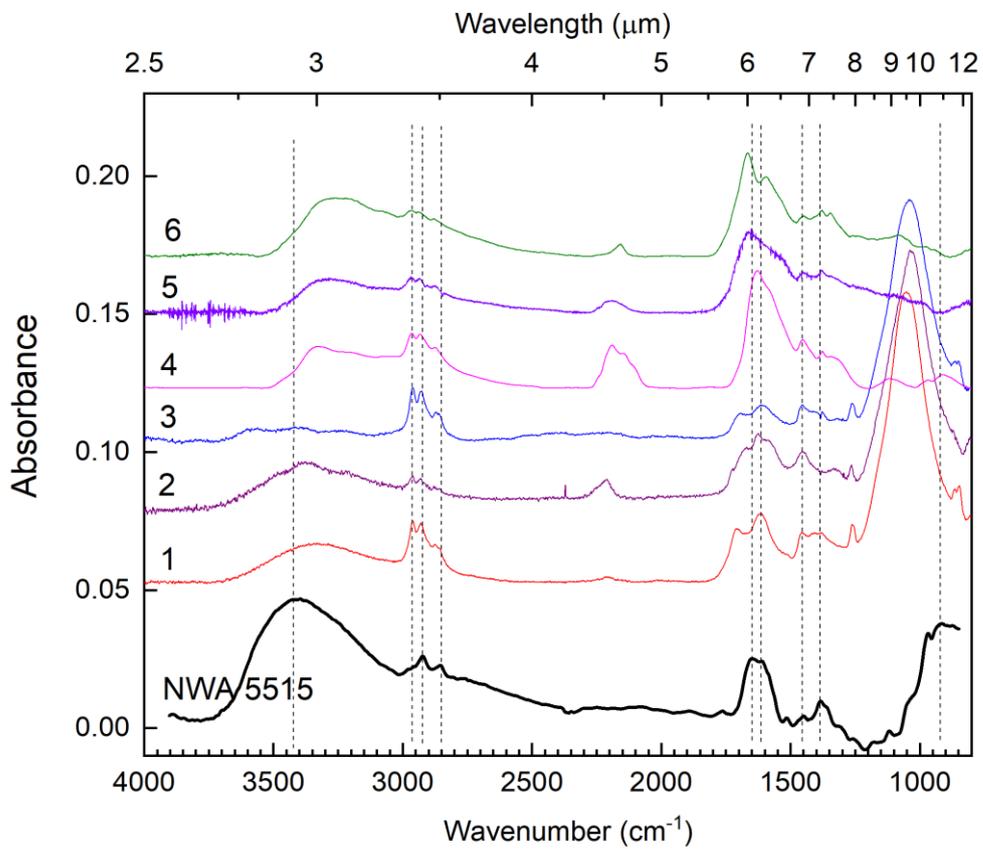

Figure 2. Comparison of the absorption IR spectra of Paris (upper) and Nord-West Africa 5515 (lower) meteorites and the laboratory organic residues produced in Jena and Catania from N-containing ice mixtures (mentioned in the figure). The spectra are vertically shifted for clarity. The vertical dashed lines indicate the most prominent absorption features in the meteorite spectra. Ice mixtures: 1 - $H_2O:CH_3OH:NH_3$ (UV irradiation, Jena), 2 - $H_2O:CO_2:NH_3$ (UV irradiation, Jena), 3 - $H_2O:CH_4:NH_3$ (UV irradiation, Jena), 4 - $N_2:CH_4$ (ion irradiation, Catania), 5 - $N_2:CH_4:CO$ (ion irradiation, Catania), 6 - $N_2:CH_4:H_2O$ (ion irradiation, Catania).

As analogs described in this paper are made from molecular ices, it is very relevant to compare their spectra with cometary ones. IDPs are derived from a larger variety of small bodies compared to macroscopic meteorites (Sandford & Bradley 1989; Bradley 2002), in particular, chondritic porous aggregate IDPs come potentially from comets (Mackinnon & Rietmeijer 1987). Thus, we used the spectra of the IDP U217B19 and IDP L2021C5 as another point of comparison to the laboratory analogs. The comparison is presented in Figure 3. The detected IR bands and their attributions for the IDPs are presented in Table 3.

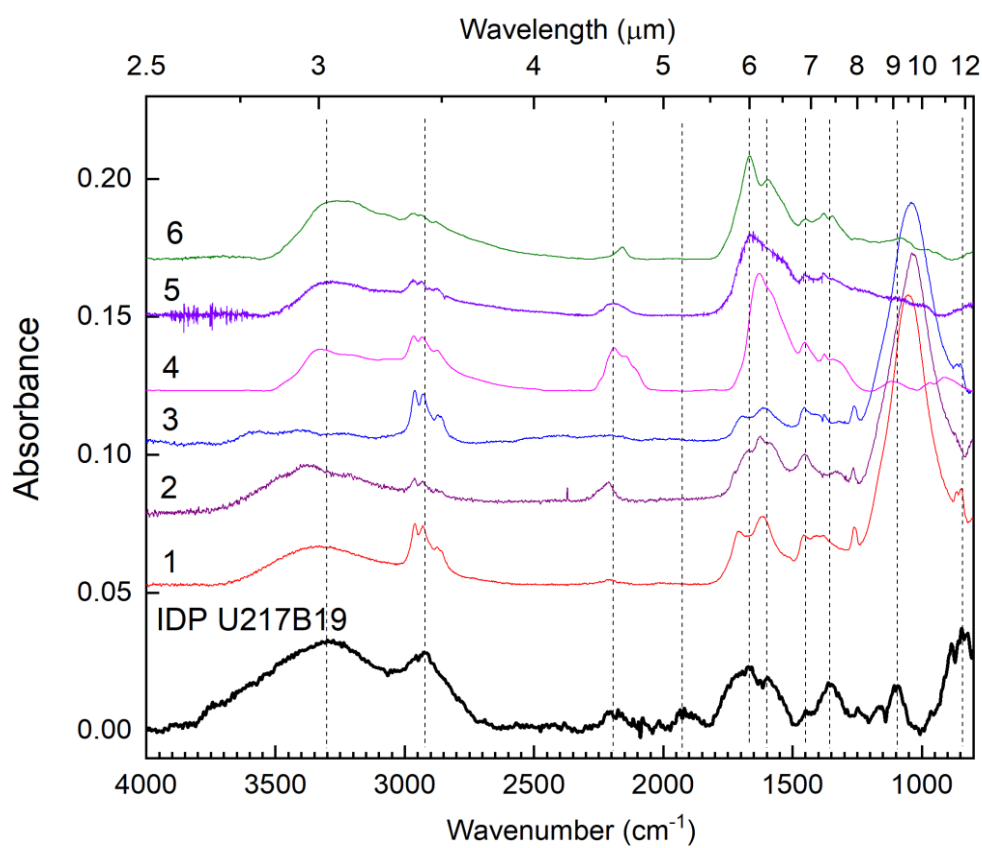

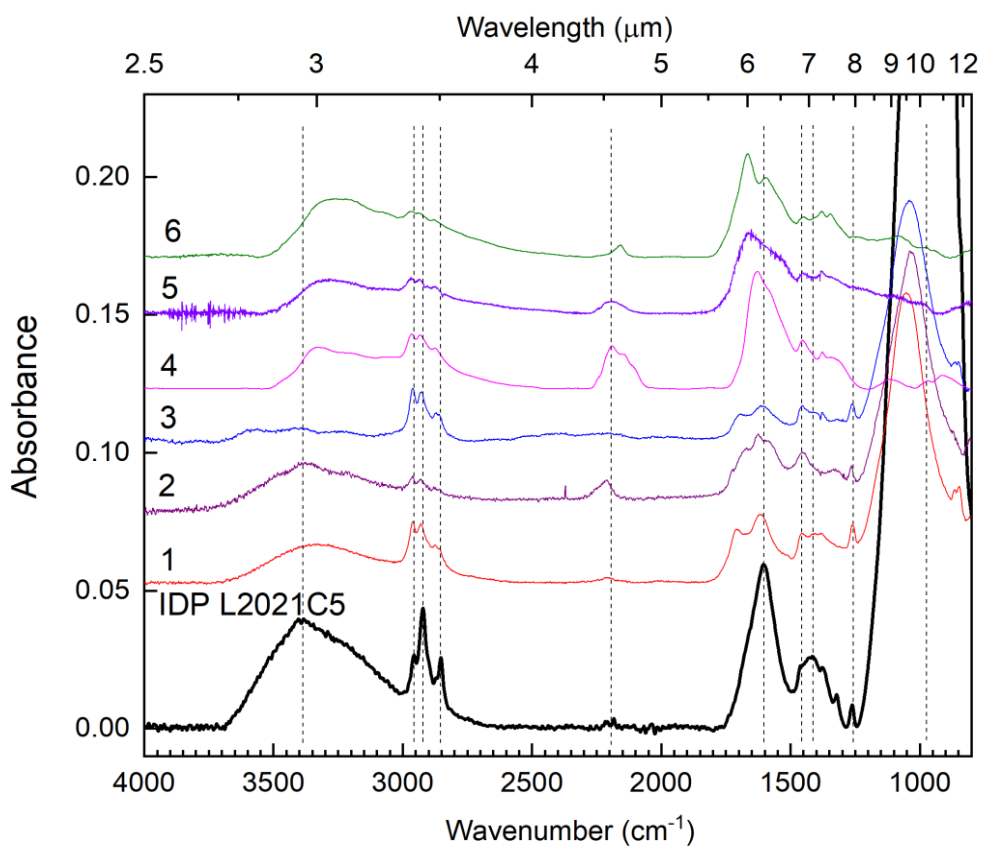

Figure 3. Comparison of the IR absorption spectra of the IDP U217B19 (upper) and IDP L2021C5 (lower) and the laboratory organic residues produced in Jena and Catania from N-containing ice mixtures (mentioned in the figure). The spectra are vertically shifted for clarity. The original spectrum of the IDP U217B19 is multiplied by 10. The vertical dashed lines indicate the most prominent absorption features in the IDPs spectra. Ice mixtures: 1 - $H_2O:CH_3OH:NH_3$ (UV irradiation, Jena), 2 - $H_2O:CO_2:NH_3$ (UV irradiation, Jena), 3 - $H_2O:CH_4:NH_3$ (UV irradiation, Jena), 4 - $N_2:CH_4$ (ion irradiation, Catania), 5 - $N_2:CH_4:CO$ (ion irradiation, Catania), 6 - $N_2:CH_4:H_2O$ (ion irradiation, Catania).

Table 3. IR bands (cm$^{-1}$) of Paris and NWA 5515 meteorites and IDPs U217B19 and L2021C5 and their attributions. The band positions are determined in this study. (p) indicates the most prominent absorption features shown in Figures 2 and 3.

| Paris meteorite | NWA meteorite | IDP U217B19 | IDP L2021C5 | Band attributions |
|---|---|---|---|---|
| 3730 – 3000 (p) | 3750 – 3000 (p) | 3800 – 3000 (p) | 3700 – 3000 (p) | OH stretching, NH stretching(Munoz Caro & Schutte 2003) |
| 2958 (p), 2920 (p) | 2963 (p), 2925 (p) | 2957, 2921 (p) | 2957 (p), 2922 (p) | CH stretching (asym)(Colangeli et al. 1995) |
| 2851 (p) | 2850 (p) | | 2870, 2854 (p) | CH stretching (sym)(Colangeli et al. 1995) |
| 2366 (p) | | | | ? |

|  |  | 2320 – 2090 (p) | 2240 – 2151 (p) | C≡N stretching(Accolla et al. 2018) |
|---|---|---|---|---|
|  |  | 1920 (p) |  | ? |
| 1662 (p), 1611 (p) | 1650 (p), 1613 (p) | 1664 (p), 1595 (p) | 1604 (p) | C=O, C=C, C=N stretching(Munoz Caro & Schutte 2003) HOH bending, $NH_2$ bending(Accolla et al. 2018) |
|  | 1515 |  | 1515 | $CO_3^{2-}$ stretching(Aguiar et al. 2009; Ciaravella et al. 2018) |
| 1454 (p) | 1449 (p) | 1450 (p) | 1461 (p) | CH bending(Munoz Caro & Schutte 2003) |
| 1418 (p) |  |  | 1415 (p) | $CO_3^{2-}$ stretching(Aguiar et al. 2009; Ciaravella et al. 2018) |
|  | 1385 (p) |  | 1378 | CH bending (sym)(Colangeli et al. 1995; Jäger et al. 2008) |
|  | 1358 | 1356 (p) |  | ? |
| 1320 | 1312 |  | 1322 | $COO^-$ stretching (sym)(Munoz Caro & Schutte 2003) |
| 1260 (p) | 1258 | 1247 | 1261 (p) | ? |
|  | 1181 | 1166 |  | ? |
|  | 1120 |  |  | ? |
| 1020 (p) | 930 (p) | 1095 (p) | 945 (p) | SiO stretching |
|  |  | 840 (p) |  | out-of-plane CH bending(Colangeli et al. 1995; Jäger et al. 2008) |

Then, we compared the spectra of the laboratory samples with those of particles directly collected in the coma of a comet. Indeed, thanks to the Stardust mission, thousands of particles with a diameter between 3 and 2000 µm were collected in the coma of the comet 81P/Wild and brought back to Earth in 2006 (Sandford 2007). Figure 4 shows the result of the comparison of the spectra of three Stardust particles and the laboratory organic residues. The detected IR bands and their attributions for the Stardust particles are presented in Table 4.

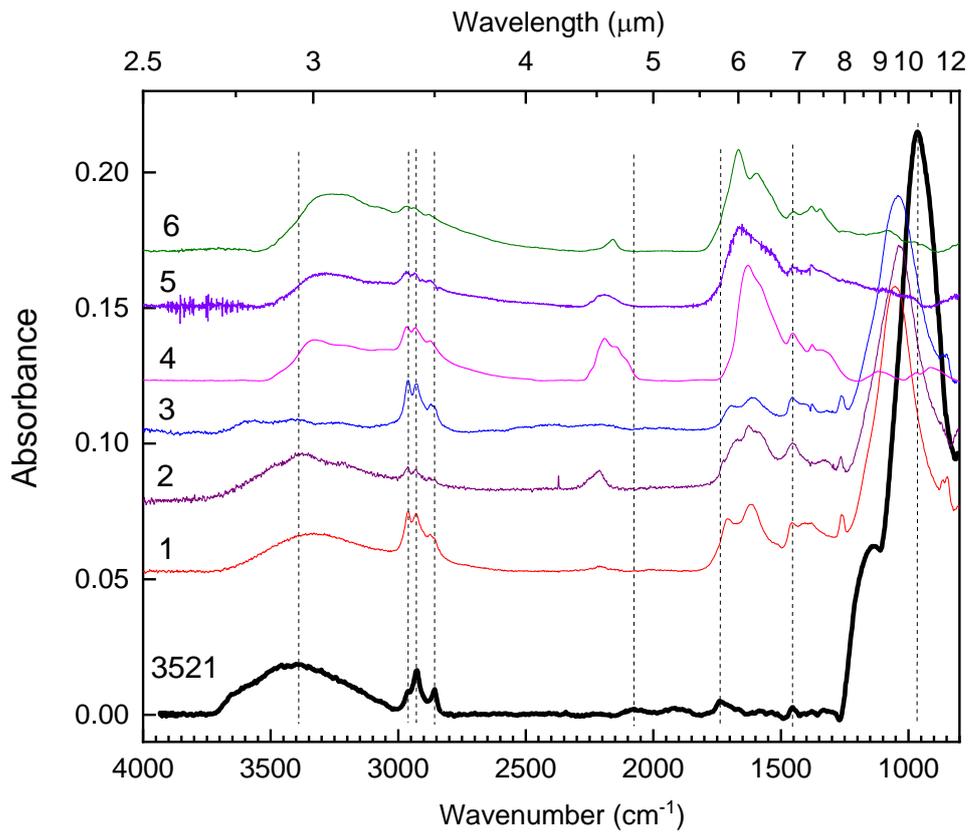
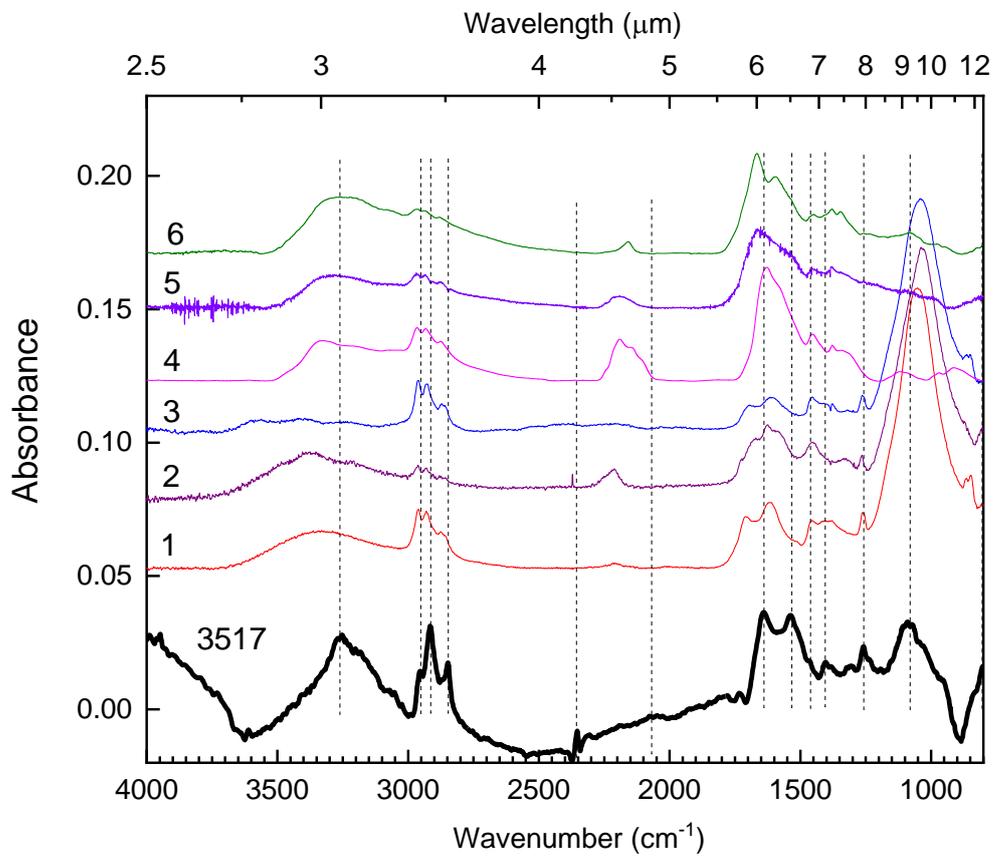

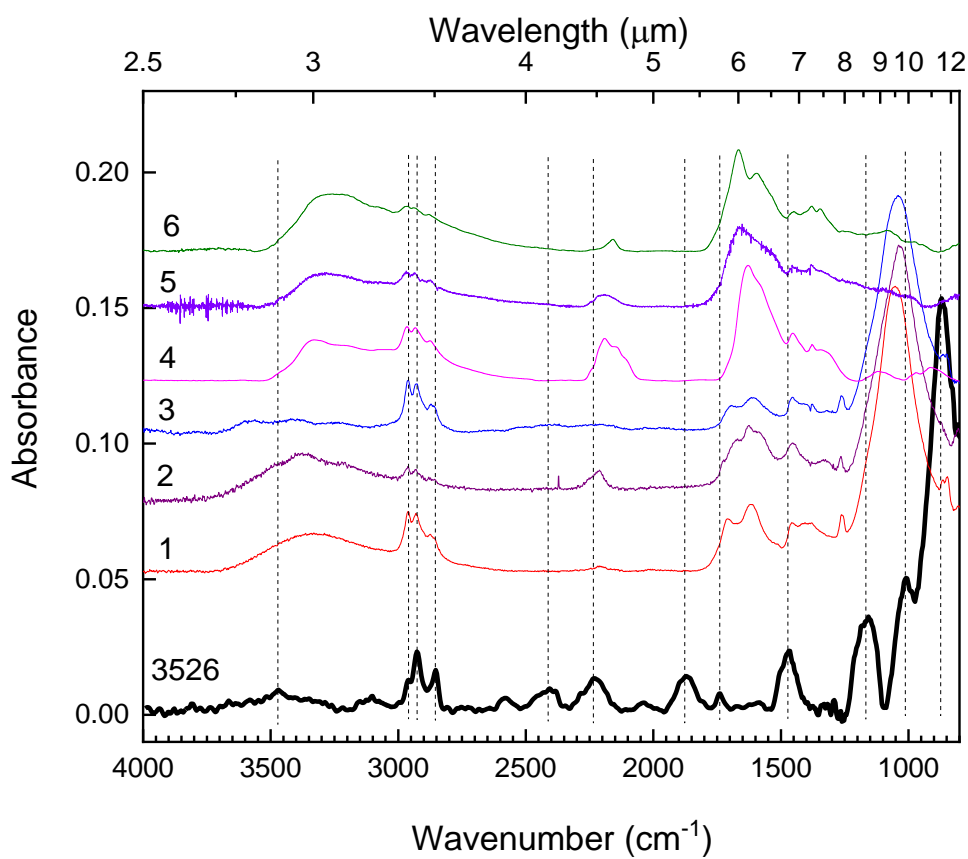

Figure 4. Comparison of the absorption spectra of the Stardust particles 3521 (upper), 3517 (middle), 3526 (lower) and the laboratory organic residues produced in Jena and Catania from N-containing ice mixtures (mentioned in the figure). The spectra are vertically shifted for clarity. The original spectra of the particles 3521 and 3526 are divided by 10. The vertical dashed lines indicate the most prominent absorption features in the particle spectra. Ice mixtures: 1 - $H_2O:CH_3OH:NH_3$ (UV irradiation, Jena), 2 - $H_2O:CO_2:NH_3$ (UV irradiation, Jena), 3 - $H_2O:CH_4:NH_3$ (UV irradiation, Jena), 4 - $N_2:CH_4$ (ion irradiation, Catania), 5 - $N_2:CH_4:CO$ (ion irradiation, Catania), 6 - $N_2:CH_4:H_2O$ (ion irradiation, Catania).

Table 4. IR bands ($cm^{-1}$) of the Stardust particles and their attributions. The band positions are determined in this study. (p) indicates the most prominent absorption features shown in Figure 4.

| 3521 | 3517 | 3526 | Band attributions |
|---|---|---|---|
| 3750 – 3000 (p) | 3600 – 3000 (p) | 3750 – 3200 (p) + 3200 - 3020 | OH stretching, NH, $NH_2$ stretching |
| 2955 (p), 3924 (p) | 2960 (p), 2925 (p) | 2960 (p), 2925 (p) | CH stretching (asym)(Colangeli et al. 1995) |
| 2854 | 2854 (p) | 2854 (p) | CH stretching (sym)(Colangeli et al. 1995) |
|  |  | 2582 | ? |
|  |  | 2410 (p) | ? |
|  | 2356 (p) |  | ? |
|  | 2310 |  | ? |
| 2150 – 2010 (p) | 2120 – 2000 (p) | 2320 – 2120 (p) | C≡N stretching (Accolla et al. 2018) |
| 1910 | | | ? |
| | | 1870 (p) | ? |
| | 1788 | | ? |
| 1739 (p), 1667, 1580 | 1730, 1638 (p) | 1740 (p), 1610, 1585 | C=O, C=C, C=N stretching (Munoz Caro & Schutte 2003) HOH bending, $NH_2$ bending (Accolla et al. 2018) |
| | 1537 (p) | | $CO_3^{2-}$ stretching (Aguiar et al. 2009; Ciaravella et al. 2018) |
| 1454 (p) | 1466 (p), 1450 | 1470 (p) | CH bending (Munoz Caro & Schutte 2003; Jäger et al. 2008) |
| | 1406 (p) | | $CO_3^{2-}$ stretching (Aguiar et al. 2009; Ciaravella et al. 2018) |
| 1378 | 1375 | | CH bending (Colangeli et al. 1995; Jäger et al. 2008) |
| | 1307 | | $COO^-$ stretching (sym) (Munoz Caro & Schutte 2003) |
| | 1260 (p) | | ? |
| | 1232 | | ? |
| | | | ? |
| 1170 | | 1160 (p) | S-O stretching |
| 962 (p) | 1084 (p), 1020, 950 | 1010 (p) | SiO stretching* |
| | | 872 (p) | out-of-plane CH bending (Colangeli et al. 1995; Jäger et al. 2008) |
| 812 | 805 (p) | | ? |

*the profile of the silicate band is due to the contribution given by the silicate in the particle as well as by the silica aerogel.

Several previous experimental investigations have focused on the formation of organic residues after UV photolysis and ion bombardment of ice mixtures (Bernstein et al. 1995; Greenberg et al. 1995; Munoz Caro & Schutte 2003; Ferini et al. 2004; Palumbo et al. 2004; Kobayashi et al. 2008; Materese et al. 2014; Baratta et al. 2015; Oba et al. 2016; Tachibana et al. 2017; Accolla et al. 2018; Urso et al. 2020). Most of these studies are based on the analysis of the samples by IR spectroscopy. This is a very powerful, non-destructive, and well consolidated analytical technique, however, it has the disadvantage that the IR bands of various functional groups occur at about the same wavelengths independent from the structure and composition of the material. To overcome this limit several studies have coupled a second technique to IR spectroscopy. As an example, high-performance liquid chromatography (Kobayashi et al. 2008; Oba et al. 2019), gas chromatography-mass spectrometry (Materese et al. 2014), optical microscopy (Tachibana et al. 2017), X-ray photoelectron spectroscopy (Accolla et al. 2018), X-ray absorption near-edge structure (Materese et al. 2014), high-resolution mass spectrometry (Oba et al. 2019), very high resolution mass spectrometry (Urso et al. 2020) have been used. The combination of different analytical techniques allows a deeper knowledge of the sample and a better comprehension of its chemical and physical properties.

As it was pointed out, relevant complex molecules in organic refractory residues have a very low abundance and have been identified by gas chromatography-mass spectroscopy, while they are difficult or even impossible to identify by IR spectroscopy (Hudson et al. 2008). In addition, by comparing the results obtained on residues produced by UV photolysis and ion irradiation of different ice mixtures, Hudson et al. (2008) suggested that energetic processing of almost any ice mixture containing C, H, N, and O atoms probably results in the formation of amino acid precursors that, if hydrolyzed, give rise to the amino acids themselves. Similar conclusions have been drawn in Materese et al. (2014).

However, IR spectroscopy remains the most suitable technique for a direct comparison with remote astronomical observations and for laboratory analysis of extraterrestrial samples. In most of the studies mentioned above the initial ice mixtures were composed of $H_2O$, $CH_3OH$, or $CH_4$, as C-bearing species, and $N_2$ or $NH_3$ as N-bearing species. In particular, $NH_3$ was used when the residue is formed after UV photolysis. In fact, as discussed in Kobayashi et al. (2008) and Islam et al. (2014) after UV photolysis of $N_2$-containing mixtures the formation efficiency of molecular species which include nitrogen is about two orders of magnitude lower than the value observed after ion irradiation. The results presented in these studies confirm that a refractory organic residue is efficiently formed after ion irradiation and UV photolysis of ice samples as far as a C-bearing species is present in the initial ice mixture. However, as pointed out in Urso et al. (2020) the detailed chemical and physical properties of the residue depend on the initial ice mixture and on the irradiation dose. These findings support the ongoing effort to perform a systematic study of the chemical and physical properties of the residues as a function of different parameters such as initial ice mixture and irradiation dose (Urso et al., in preparation).

All extraterrestrial particles of our study (meteorites, IDPs, and cometary particles) present mixtures of organic components and minerals, and both have proper signatures in the IR spectra. The silicate features are the most prominent in the spectra. As one can see from the comparisons presented in Figures 2 - 4 and from Tables 1 - 4, the majority of the IR bands caused by the presence of organic materials in the studied particles can be reproduced in the spectra of the laboratory analogs. However, the spectra of the particles and the spectra of the residues do not match perfectly. This is not surprising as we consider a very limited number of ice mixtures with only a few molecular species and with particular abundance ratios. On one hand, none of these ice mixtures is exactly equal to the expected chemical composition of ice grain mantles or ice surfaces of Solar System bodies. On the other hand, we have considered mixtures made of molecules that have been observed or are expected to be present in astrophysical ices. Quantitative differences between the laboratory spectra clearly defined by the initial ice mixture, type, energy, and conditions of the processing are observed and, if studied systematically, will clarify the genesis of the extraterrestrial organic materials. Below, we discuss a few important results, which have been obtained in our study.

The broad band is observed in the spectra of all extraterrestrial particles in the range of 3750 – 3000 cm$^{-1}$ (2.67 – 3.33 μm). For the Stardust particle 3517 the range is a bit narrower, 3600 – 3000 cm$^{-1}$ (2.78 – 3.33 μm), however, it can be a baseline problem. For the Stardust particle 3526 the band is split into two. The band is observed in the same range in the spectra of the Jena samples. It has been previously shown by the Jena group that some water molecules mixed with silicates at low temperatures are trapped on silicate grains beyond the desorption temperature of water ice (Potapov et al. 2018a; Potapov et al. 2018b; Potapov et al. 2021). Trapped water presenting, probably, water molecules strongly bound in hydrophilic binding sites on silicate grains shows up as the weak OH-stretching and OH-bending bands in the 3600 and 1600 cm$^{-1}$ (2.78 and 6.25 μm) spectral regions correspondingly as is observed in this study for the Jena samples. For the Catania samples, the band is narrower, 3500 – 3000 cm$^{-1}$ (2.86 – 3.33 μm). As far as it was also detected in "oxygen-less" samples, it was mainly referred to the –NH$_2$ and –NH–groups and, additionally, to the –OH functional groups in the case of O-containing ices (Accolla et al. 2018). Comparing the two sets of samples, Jena and Catania, we can clearly attribute the spectral range 3750 – 3500 cm$^{-1}$ (2.67 – 2.86 μm) to the OH stretching of water molecules trapped on silicate grains and the spectral range 3500 – 3000 cm$^{-1}$ (2.86 – 3.33 μm) to both NH and OH groups in the residue. Note that a contribution from OH-groups in alcohols is also possible in these spectral ranges.

The band located between 2300 and 2000 cm$^{-1}$ (4.35 and 5.0 μm) related to the C≡N triple bond is detected in the spectra of both IDPs and of the Stardust particles 3521 and 3526 and is not detected in the meteorites. The detection of this feature in extraterrestrial particles is an important result meaning that N-bearing species were present in the particle material at the formation of the organic residue.

In the spectra of the extraterrestrial particles (both meteorites, IDP L2021C5, and Stardust particle 3517), we see evidence of the presence of organic carbonates with IR signatures between 1410 - 1430 and 1520 - 1540 cm$^{-1}$ (~ 7.0 and 6.5 μm). Furthermore, a set of organics signatures is present in the spectra of the extraterrestrial particles and their analogs: the CH$_2$ and CH$_3$ aliphatic stretching bands, localized respectively around 2930 cm$^{-1}$ and 2960 cm$^{-1}$ (3.41 and 3.38 μm) for the asymmetric vibration and 2850 cm$^{-1}$ and 2870 cm$^{-1}$ (3.51 and 3.48 μm) for the symmetric vibration; the C=O stretching band (around 1700 cm$^{-1}$, 5.88 μm), C=C and C=N stretching bands (around 1630 and 1580 cm$^{-1}$, 6.13 and 6.33 μm), CH bending bands

(around 1460, 1380, 1260, 850, 760 and 660 cm$^{-1}$; 6.85, 7.25, 7.94, 11.76 and 15.15 μm). However, their attributions to definite species (if possible) need an additional structural and elementary analysis, which is out of the scope of the present study.

Most of the extra-terrestrial samples presented in this study (Paris and NWA 5515 meteorite, IDP L2021C5, and Stardust particle 3521) show a broad silicate band with a maximum between 930 and 1045 cm$^{-1}$ (10.75 and 9.57 μm). The variation of the maximum could be due to compositional or structural differences. For instance, primitive meteorites, such as Paris, showing a matrix with partly amorphous hydrated silicates, will present a maximum above 1000 cm$^{-1}$ (10 μm). On the contrary, NWA 5515 is a more heated meteorite and exhibits a spectral signature around 930 cm$^{-1}$ (10.75 μm), which corresponds to a matrix rich in anhydrous phases. Conversely, some samples exhibit a silicate signature composed of several peaks, such as particle 3526. This sample is mainly composed of crystalline olivine (bands at 1017 and 871 cm$^{-1}$, 9.83 and 11.48 μm) (Rotundi et al. 2008). Indeed, complex structures of the silicate band in the spectra of the extra-terrestrial particles suggest crystalline silicates. In the cases of the Paris meteorite and the IDP L2021C5 presenting more amorphous silicates, we observe a good coincidence with the silicate band in the spectra of the Jena samples. In the case of the Stardust particles, the profile of the silicate band is due to the contribution given by the silicate in the particle as well as by the silica aerogel, in which each particle has been embedded. Thus, in this case, the comparison with laboratory analogs is not straightforward.

## 4. Discussion
### 4.1. Pathways to refractory organics

Similarities between the spectra of the laboratory samples and the extraterrestrial particles reinforce the idea of a pathway for the formation of refractory organics through energetic and thermal processing of molecular ices in the solar nebula or in its parent molecular cloud. Ion irradiation and UV photolysis of ice mixtures could significantly contribute to the formation of the organic matter observed in extraterrestrial particles. Of course, this does not exclude that other processes were at work as well. The laboratory spectra of the samples produced by UV or ion processing are qualitatively very similar to each other. This allows us to make the conclusion that the formation of an organic residue is a general result independent of the initial mixture (as far as C-bearing species are included) and of the type of energetic processing. The same conclusion was reached in a number of previous studies (Islam et al. 2014; Munoz Caro et al. 2014). Slight differences in the composition and structure of the residues depend on the initial mixture, the type of energetic processing, and the photon or ion fluence. This was also demonstrated in studies of organic residues of different ice mixtures after VUV processing (Bernstein et al. 1995; Greenberg et al. 1995).

Below, we compare the ion and UV fluxes and doses in different astrophysical environments (see also Table 5) and their possible effects on interstellar and circumstellar ices. It has been estimated that in dense molecular clouds the effective 1 MeV proton flux of cosmic ions is 1 cm$^{-2}$ s$^{-1}$ (Mennella et al. 2003) and the UV flux, due to the cosmic-ray-induced emission of molecular hydrogen, is of the order of $2 \times 10^3 - 3 \times 10^4$ photons cm$^{-2}$ s$^{-1}$, depending on the assumed cosmic-ray spectrum at low energy (Prasad & Tarafdar 1983; Shen et al. 2004). As estimated by Shen et al. (2004), after $10^7$ years the dose deposited by cosmic rays on water-ice grain mantles varies in the range 1-10 eV/molecule, and the dose deposited by UV photons varies in the range 10-100 eV/molecule. Based on these results, the energy input by UV photons is about an order of magnitude higher than the energy input by cosmic-ray particles.

Moore (1999) showed that in diffuse regions of the ISM UV photons deposit more energy in icy mantles than 1 MeV protons. In particular, after $10^7$ years the dose deposited by UV photons is of the order of $10^6$ eV/molecule and the dose deposited by low-energy cosmic rays is on the order of 30 eV/molecule. It should be noted that icy grain mantles are not observed in diffuse regions. This is ascribed to the fact that the photon-induced desorption rate is higher than the gas adsorption rate (Westley et al. 1995). However, evidence of the presence of solid-state water (similar to the trapped water discussed in this study) in the diffuse ISM has been provided recently by Potapov et al. (2021) on the basis of the combination of laboratory data and infrared observations.

Pedersen & Gómez De Castro (2011) have modeled the protoplanetary disk of a young stellar object assuming that the UV radiation field of a T Tauri star at 500 AU from the central object is $2.9 \times 10^{10}$ photons cm$^{-2}$ s$^{-1}$. Strazzulla et al. (1983) have estimated that the flux of 1 MeV protons in a T Tauri star is $2 \times 10^{10}$ cm$^{-2}$ s$^{-1}$ at 1 AU. Assuming that the flux decreases with distance as d$^{-2}$, at 500 AU the value is $8 \times 10^4$ cm$^{-2}$ s$^{-1}$.

Speaking about the Solar System, it has been estimated that near the cometary surface in Oort cloud the dose deposited by cosmic rays after $10^9$ years is as high as hundreds of eV/16u, where u is the unified atomic mass unit defined as 1/12 of the mass of an isolated atom of carbon-12 (Strazzulla & Johnson 1991; Modica et al. 2012). Stellar UV fluxes for Sun-like stars have been estimated in the range from $10^{11}$ photons cm$^{-2}$ s$^{-1}$ at 1 AU to $10^7$ photons cm$^{-2}$ s$^{-1}$ at 100 AU (Torsten Löhne, private communication). Assuming that the flux decreases with distance as d$^{-2}$, this gives us $10^5 - 10^3$ photons cm$^{-2}$ s$^{-1}$ at $10^3 - 10^4$ AU.

Table 5. Comparison of the ion and UV fluxes and doses in different astrophysical environments.

| Environment | Time | Energetic ions | | UV photons | | | Reference |
|---|---|---|---|---|---|---|---|
| | years | Flux (ions cm$^{-2}$ s$^{-1}$) | Dose (eV/molecule) | Flux (photons cm$^{-2}$ s$^{-1}$) | Dose (eV/molecule) | Fluence (photons cm$^{-2}$) | |
| Diffuse interstellar clouds | $10^7$ | 10 | 30 | $9.6\times10^7$ | $10^6$ | $3\times10^{22}$ | (Moore 1999) |
| Dense molecular clouds | $10^7$ | 1 | 1-10 | $2\times10^3 - 3\times10^4$ | 10-100 | $6\times10^{17}$ - $9\times10^{18}$ | (Shen et al. 2004) |
| T-Tauri disks (500 AU) | $10^6$ | $8\times10^4$ | $2.4\times10^4$ | $2.9\times10^{10}$ | $2.9\times10^7$ | $9\times10^{23}$ | (Strazzulla et al. 1983; Pedersen & Gomez de Castro 2011) |
| Cometary nuclei (Oort cloud) | $10^9$ | 10 | 600 (down to a depth of a few meters) | $10^3$–$10^5$ | $10^3$–$10^5$ (down to a depth of a few 100s nm) | $3\times10^{19}$ - $3\times10^{21}$ | (Strazzulla & Johnson 1991; Modica et al. 2012) |

Thus, the dose of UV photons irradiating ice grain mantles in the ISM and planet-forming disks is orders of magnitude higher than that of energetic ions. On the other hand, UV photons are absorbed in the outermost layers, on the order of $10^2$ nm (Cruz-Diaz et al. 2014a, 2014b), depending on the optical constants of the target material (Baratta et al. 2002). Thus, their effects would be important for small grains and surface processes and would be negligible for large bodies with respect to the effects caused by energetic ions which can penetrate as deep as a few meters (Strazzulla & Johnson 1991; Modica et al. 2012). In addition, the effects of UV photons would be important if $NH_3$ (and not $N_2$) is the main N-bearing species in icy grain mantles. If $N_2$ is the main N-bearing species in interstellar and circumstellar ices, UV photons compared to ions would produce negligible chemical effects due to the very low absorption of UV light by $N_2$ (Hudson & Moore 2002; Kobayashi et al. 2008; Wu et al. 2012; Islam et al. 2014).

Ion irradiation experiments have been intentionally performed in order to reach a dose of the order of 100 eV/16u because this dose is able to produce a refractory residue and it is relevant for comets and TNOs (Strazzulla & Johnson 1991; Baratta et al. 2019). The fluence used in UV irradiation experiments ($2\times10^{19}$ photons cm$^{-2}$) is of the same order as the highest fluence suffered by icy grain mantles in dense molecular clouds. In this view, our experiments show that the refractory organic material observed in meteorites, IDPs, and cometary dust particles could have been formed either in the protostellar phase or in the Solar System.

### 4.2. Formation location of the extraterrestrial particles

Due to the spectral signatures of the meteorites, IDPs, and Stardust particles in the region of 3750 – 3000 cm$^{-1}$, where trapped water is detected in the Jena samples, we conclude that trapped water is probably present in the extraterrestrial particles. Note, that trapped water originates from ice/silicate mixtures at low temperatures and its formation does not require high-temperature processing as in the case of hydrated silicates (phyllosilicates) where OH groups or H$_2$O molecules are chemically incorporated into crystalline silicates. Thus, the formation of trapped water is relevant to both meteorites and comets as formed in low-temperature regions beyond the water snowline.

This conclusion is reinforced by the detection of the C≡N triple bond in the IDPs and Stardust particles meaning that N-bearing species were present in the particle material at the formation of the organic residue. This points out that comets have originated in the outer regions of the solar nebula, beyond the snowline of one of the abundant in cometary ices N-bearing species, such as nitrogen (snowline at about 40 K) or ammonia (snowline at about 100 K). The non-detection of the C≡N bond in the meteorite spectra (see Figure 2 and Table 3) reinforces this conclusion as meteorites are derived from asteroids formed in the region between 2 and 4 AU, at temperatures much higher than 100 K. This result is in line with the finding of a nitrogen-rich organic matter in ultracarbonaceous micrometeorites recovered from Antarctica (representing a small fraction of interplanetary dust particles reaching the Earth's surface), which was also referred to the formation location of the organic matter in low-temperature regions of the Solar System (Dartois et al. 2013; Dartois et al. 2018).

Recent literature on comets agrees with the fact that comets' organics come from the interplanetary medium (Raponi et al. 2020). However, we want to note that several authors proposed that organic compounds could be formed in the envelopes of evolved stars and delivered through the ISM to our environment (Papoular 2001; Kwok & Zhang 2011; Endo et al. 2021). The formation of –CN groups (nitrile or cyanide groups) is observed in gas-phase condensation processes performed in molecular nitrogen. It is generally known that the condensation of carbon particles in a nitrogen atmosphere forms a big amount of nitriles in the carbon structure. The incorporation of nitrogen (in the form of nitrile groups) into the refractory carbonaceous matter is efficient. The use of NH$_3$ is less efficient for nitrile formation (Alexandrescu et al. 1998; Thareja et al. 2002). Consequently, circumstellar envelopes around late-type stars can produce carbonaceous grains containing nitriles. A top-down synthesis of carbonaceous matter or molecules containing nitriles or other nitrogen-bearing molecules from circumstellar dust could be possible. However, this idea needs further investigation. For now, there is no convincing evidence for such a scenario.

Detection of vibrational signatures of carbonate groups related to amorphous Mg-carbonate in combination with the formation of the extraterrestrial particles beyond the nitrogen or ammonia snowline speaks for their formation in cold regions (outer solar nebula or its parent ISM). To clarify the origin of carbonates, they should be searched for in diffuse and dense interstellar clouds. The James Webb Space Telescope (JWST) may provide a sensitivity high enough for the detection of carbonates in the ISM.

### 4.3. Origin of water on Earth and terrestrial planets

Detection of trapped water in extraterrestrial particles (some of which have reached Earth) is linked to another intriguing question – could the building blocks of Earth be wet? Assuming that solid-state water exists only beyond the water snowline in planet-forming disks, two scenarios for the delivery of water to the surface of terrestrial planets have been proposed (van Dishoeck et al. 2014). In the so-called dry scenario, the planets are initially built up from planetesimals/pebbles inside the snowline with low (or no) water mass fractions, and water is delivered to their surfaces by water-rich asteroids formed outside the snowline. Alternatively, in the so-called wet scenario, the planets either accreted a water-rich atmosphere or formed beyond the snowline. For instance, in the pebble accretion scenario, the pebbles that form planets drift inward from the outer disk regions, carrying water ice with them. The ice evaporates resulting in water vapor diffusing into the inner disk (Bitsch et al. 2021).

However, one more possibility of the wet scenario is that local planetesimals/pebbles in the time of the Earth formation retained some water at high temperatures through its physisorption or chemisorption on silicate grains. This is exactly the phenomenon, which we observe in the laboratory – trapped water on silicates. In the present study, we provide evidence for the presence of trapped water in comets and meteorites. Such objects could deliver water to the early Earth.

A search for water trapped on silicates in the inner regions (inside the $H_2O$ snowline) of planet-forming disks should be also possible with JWST and may shed light on the origin of water on Earth and terrestrial planets (and, potentially, extrasolar planets). Appropriate observations are on the horizon.

**Acknowledgments**

We thank Hope Ishii for providing the IDP U217B19 data and Torsten Löhne for providing the values for stellar UV fluxes. AP and CJ acknowledge support by the Research Unit FOR 2285 "Debris Disks in Planetary Systems" of the Deutsche Forschungsgemeinschaft (grant JA 2107/3-2). MEP acknowledges the support by the Italian Ministero dell'Istruzione, dell'Università e della Ricerca through the grant Progetti Premiali 2012-iALMA. GAB acknowledges the financial supported from the Italian Space Agency (ASI) contract no. 2013-073-R.0: PSS (Photochemistry on the Space Station). TH acknowledges support from the European Research Council under the Horizon 2020 Framework Program via the ERC Advanced Grant Origins 83 24 28.